\documentclass[sn-basic]{sn-jnl}

\usepackage{graphicx}%
\usepackage{multirow}%
\usepackage{amsmath,amssymb,amsfonts}%
\usepackage{amsthm}%
\usepackage{mathrsfs}%
\usepackage[title]{appendix}%
\usepackage{xcolor}%
\usepackage{textcomp}%
\usepackage{manyfoot}%
\usepackage{booktabs}%
\usepackage{algorithm}%
\usepackage{algorithmicx}%
\usepackage{algpseudocode}%
\usepackage{listings}%
\usepackage[normalem]{ulem}
\usepackage{float}
\usepackage{adjustbox}
\usepackage{graphicx}
\usepackage{caption}
\usepackage{multirow}
\usepackage{booktabs}
\usepackage{graphicx}
\usepackage{lineno}
\usepackage{tabularx}

\begin{document}

\newcommand{\yan}[1] {\textcolor{red}{#1}} 
\newcommand{\del}[1] {\textcolor{red}{\st{#1}}}

\title[MRISegmentator-Abdomen: A Fully Automated Multi-Organ and Structure Segmentation Tool for T1-weighted Abdominal MRI]{MRISegmentator-Abdomen: A Fully Automated Multi-Organ and Structure Segmentation Tool for T1-weighted Abdominal MRI}



\author*[1]{\fnm{Yan} \sur{Zhuang}}\email{yan.zhuang2@nih.gov}

\author[1]{\fnm{Tejas Sudharshan} \sur{Mathai}}\email{tejas.mathai@nih.gov}
\equalcont{These authors contributed equally to this work.}

\author[1]{\fnm{Pritam} \sur{Mukherjee}}\email{pritam.mukherjee@nih.gov}
\equalcont{These authors contributed equally to this work.}

\author[2]{\fnm{Brandon} \sur{Khoury}}\email{brandon.r.khoury.mil@health.mil}

\author[1]{\fnm{Boah} \sur{Kim}}\email{boah.kim@nih.gov}

\author[1]{\fnm{Benjamin} \sur{Hou}}\email{benjamin.hou@nih.gov}

\author[3]{\fnm{Nusrat} \sur{Rabbee}}\email{nusrat.rabbee@nih.gov}

\author[1]{\fnm{Abhinav} \sur{Suri}}\email{rsummers@cc.nih.gov}
 
\author[1]{\fnm{Ronald M.} \sur{Summers}}\email{rsummers@cc.nih.gov}

\affil*[1]{\orgdiv{Radiology and Imaging Sciences}, \orgname{National Institutes of Health Clinical Center}, \orgaddress{\street{10 Center Dr}, \city{Bethesda}, \postcode{20892}, \state{MD}, \country{USA}}}

\affil[2]{\orgdiv{Department of Radiology}, \orgname{Walter Reed National Military Medical Center}, \orgaddress{\street{8901 Rockville Pike}, \city{Bethesda}, \postcode{20892}, \state{MD}, \country{USA}}}

\affil[3]{\orgdiv{Biostatistics and Clinical Epidemiology Services}, \orgname{National Institutes of Health Clinical Center}, \orgaddress{\street{10 Center Dr}, \city{Bethesda}, \postcode{20892}, \state{MD}, \country{USA}}}


\abstract{
\textbf{Background} Segmentation of organs and structures in abdominal MRI is useful for many clinical applications, such as disease diagnosis and radiotherapy. Current approaches have focused on delineating a limited set of abdominal structures (13 types). To date, there is no publicly available abdominal MRI dataset with voxel-level annotations of multiple organs and structures. Consequently, a segmentation tool for multi-structure segmentation is also unavailable.

\noindent
\textbf{Methods} We curated a T1-weighted (T1w) abdominal MRI dataset consisting of 195 patients who underwent imaging at National Institutes of Health (NIH) Clinical Center. The dataset comprises of axial pre-contrast T1, arterial, venous, and delayed phases for each patient, thereby amounting to a total of 780 series (69,248 2D slices). Each series contains voxel-level annotations of 62 abdominal organs and structures. A 3D segmentation (nnUnet) model, dubbed as MRISegmentator-Abdomen (MRISegmentator in short), was trained on this dataset, and evaluation was conducted on an internal test set and two large external datasets: AMOS22 and Duke Liver. The predicted segmentations were compared against the ground-truth using the Dice Similarity Coefficient (DSC) and Normalized Surface Distance (NSD). 

\noindent
\textbf{Findings} MRISegmentator achieved an average DSC of 0.861±0.170 and a NSD of 0.924±0.163 in the internal test set. On the AMOS22 dataset, MRISegmentator attained an average DSC of 0.829±0.133 and a NSD of 0.908±0.067. For the Duke Liver dataset, an average DSC of 0.933±0.015 and a NSD of 0.929±0.021 was obtained. 

\noindent
\textbf{Interpretation} The proposed MRISegmentator provides automatic, accurate, and robust segmentations of 62 organs and structures in T1-weighted abdominal MRI sequences. The tool has the potential to accelerate research on various clinical topics, such as abnormality detection, radiotherapy, disease classification among others. The segmentation model is publicly available at \url{https://github.com/rsummers11/MRISegmentator} \footnote{The dataset will be publicly available upon acceptance of the paper.}. 


\noindent
\textbf{Funding} This work was supported by the Intramural Research Program of the National Institutes of Health (NIH) Clinical Center (project number 1Z01CL040004)
}




\maketitle

\section{Introduction}

Magnetic Resonance Imaging (MRI) is a fundamental and versatile tool in medical imaging, and it is widely incorporated into clinical workflows~(\cite{Dirix2014_valueOfMRI}) due to its comprehensive diagnostic capabilities. 
Automating the delineation of organs and structures on MRI sequences can enable downstream applications, such as the early detection of cancer (\cite{mathai2024_detectionAbdPelvicLN}), tracking interval changes in size of radiologic findings, the diagnosis of diffuse and focal liver disease (\cite{macdonald2023duke}), radiotherapy planning and guidance (\cite{keall2022integrated}), and quantification of body composition for opportunistic screening of diseases (\cite{zaffina2022body}). Auto-segmentation of multiple structures can improve patient outcomes without increasing the burden on radiologists (\cite{zhu2023utilizing}).

Numerous studies have explored MRI segmentation across various body parts, such as the brain, heart, abdomen, and pelvis~(\cite{billot2023robust,zhuang2019evaluation, macdonald2023duke,nyholm2018mr}). Despite recent advancements, these works focus on a limited set of organs and structures. In particular, multi-organ and structure segmentation for abdominal MRI lags significantly behind its CT counterpart. Furthermore, the process of annotating multiple organs and structures in radiologic studies is not only time-consuming and labor-intensive, but often requires medical expertise that can make the annotation process prohibitively expensive~(\cite{greenspan2016guest}). This challenge is amplified in the case of the abdomen, which is an anatomically complex area with several subtle structures. As a result, there is a noticeable lack of large-scale, high-quality MRI datasets with voxel-level annotation, detailed information on MRI acquisition, and transparent data on patient demographics. 


Presently, the AMOS22 dataset~(\cite{ji2022amos}) stands as the largest abdominal MRI dataset that is publicly available to the best of our knowledge. However, annotations were only provided for 13 key abdominal organs across 60 patients, and comprehensive data acquisition and patient demographics information were not available. Meanwhile, the Duke Liver MRI dataset, which was recently released, restricted the annotations provided solely to the liver~(\cite{macdonald2023duke}). In contrast, there are several publicly available abdominal CT datasets~(\cite{Ma-2021-AbdomenCT-1K}) that contain the requisite amount of data for training robust automated tools. These datasets have contributed to the development of a powerful CT-based multi-structure segmentation tool, TotalSegmentator (TS), that can delineate 117 different structures in the body~(\cite{wasserthal2023totalsegmentator}). It is the current \textit{de-facto} standard for benchmarking multi-organ and structure segmentation tasks in CT. However, such a tool for abdominal MRI does not exist currently, and there is a pressing demand to create a robust and accurate tool for abdominal MRI.

\begin{figure}[!htb]
    \centering
    \begin{minipage}{1\textwidth}
        \centering
        \includegraphics[width=1\linewidth]{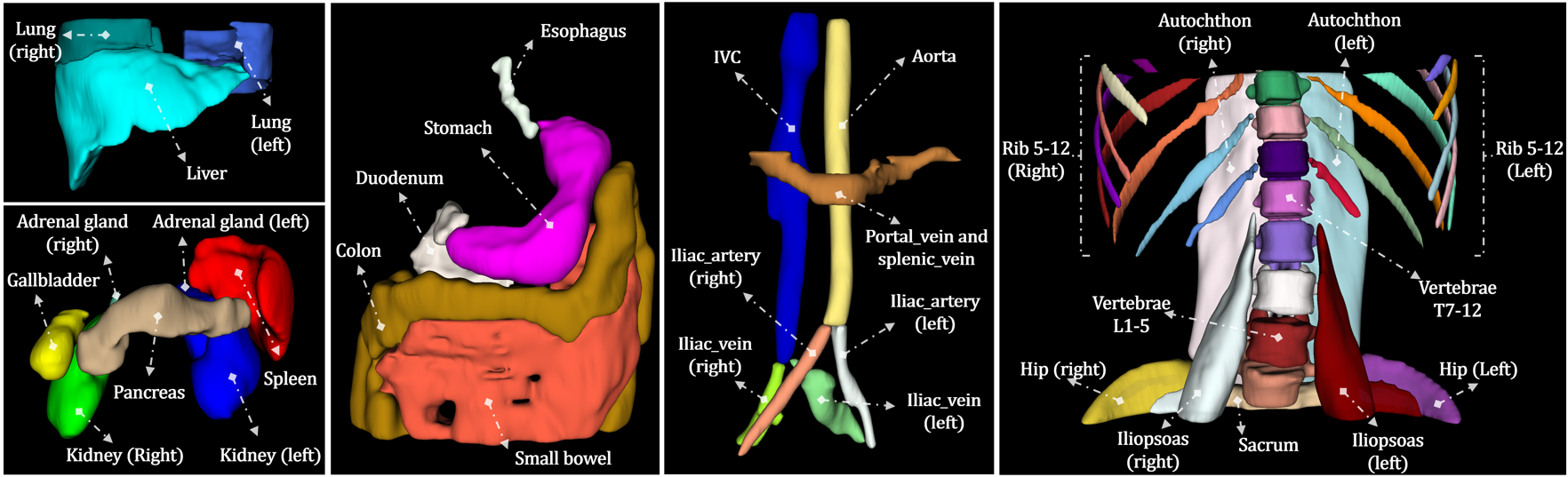} 
    \end{minipage}
\caption{We developed a multi-organ and structure segmentation tool called MRISegmentator-Abdomen (\textbf{MRISegmentator} in short) to segment 62 organs and structures in abdominal T1w MRI studies. The tool comprehensively covers various anatomical regions in the abdomen including 15 major organs, 7 vessels, 8 muscles, and 32 skeletal structures. Visual examples of the segmentations for these organs and structures are shown above.}
\label{Fig:list_of_organ} 
\end{figure}

In this study, we curated a large-scale multi-parametric abdominal T1-weighted (T1w) MRI dataset. The dataset contains 780 volumes, including both pre-contrast T1-weighted (T1w PRE) and contrast-enhanced T1-weighted in the arterial (T1w ART), portal venous (T1w VEN), and delayed (T1w DEL) phases, totalling 69,248 2D slices from 195 unique patients. As shown in Fig.~\ref{Fig:list_of_organ}, voxel-level annotations for 62 organs and structures across various regions in the abdomen were obtained. The annotation process was comprised of a two-step semi-automated annotation pipeline. First, cross-modality segmentation methods were utilized for producing large-scale pseudo-labels to reduce annotation time. Then, an iterative learning strategy was adopted, in which successive models were trained on refined versions of annotations derived from the outputs of previous models. This led to powerful models being progressively created with improved annotations obtained in each round. Once the annotations were available for all structures in all volumes, the proposed MRISegmentator-Abdomen (\textbf{MRISegmentator} in short) model was developed to segment structures in abdominal T1w MRI. Evaluations on an internal dataset showed that the tool effectively segmented 15 major organs, 7 blood vessels, 8 muscles groups, and 32 bones  within the abdomen. Furthermore, external validation on the AMOS22 and Duke Liver datasets corroborated its robustness to studies acquired at external institutions. We make the MRISegmentator model publicly available at \url{https://github.com/rsummers11/MRISegmentator}.\footnote{The dataset will be publicly available upon acceptance of the paper.}


\section{Materials and Methods}

\subsection{Patient Sample}

This retrospective study utilized three different datasets, one internal  and two external, which are described next. 

\medskip
\noindent
\textbf{Internal Dataset.} The internal dataset was Health Insurance Portability and Accountability Act (HIPAA) compliant and approved by the Institutional Review Board at NIH Clinical Center. The requirement for signed informed consent from the patients was waived. The Picture Archiving and Communication System (PACS) system at the National Institutes of Health Clinical Center (NIH CC) was queried for patients who underwent both abdominal MRI and CT scans on the same day, between January 2019 and October 2021 at the NIH CC~(\cite{mathai2023universal, mathai2022lymph}). As shown in Fig.~\ref{Fig:inclusion_exclusion_internal} in supplementary materials, out of 632 patients who met the search criteria, 225 were randomly selected for the annotation. The patients in the sample had a variety of underlying pathological conditions, such as liver tumors and pancreatic cysts. 30 patients were excluded as they had severe metastatic disease and annotation of all the lesions for these patients would be very cumbersome. As a result, the final patient sample consisted of 195 patients.

Each patient had all four axial T1 images: (1) pre-contrast T1-weighted (T1w PRE), (2) contrast-enhanced T1-weighted in the arterial (T1w ART), (3) portal venous (T1w VEN), and (4) delayed (T1w DEL) phases, respectively. This resulted in a total of 780 MRI images for 195 patients. CT images acquired on the same day for these 195 patients were also collected, and these images contained both contrast and non-contrast images. Table~\ref{tab-internal-metaInfo} in the supplementary materials summarizes the acquisition parameters and demographic details of the internal dataset. The CT images served as a base to generate pseudo-annotations for the MRI sequences, which will be discussed in Section~\ref{sec:anno}. Their acquisition information is shown in Table~\ref{tab:ct_info} in the supplementary materials.

\medskip
\noindent
\textbf{External Dataset.} The first of two external datasets was a subset of the MRI data released as part of the Multi-Modality Abdominal Multi-Organ Segmentation Challenge 2022 (AMOS22) challenge~(\cite{ji2022amos}). The AMOS22 dataset (MRI subset) contained 60 multi-planar, multi-sequence MRI volumes from 60 patients. It provided segmentation annotations for 13 main abdominal organs, including the kidneys, liver, and stomach, among others. However, it did not provide detailed information related to the sequence type and patient demographics. Thus, the AMOS22 dataset (whole MRI subset) was employed to evaluate the segmentation performance of MRISegmentator on these 13 major organs.

The second external dataset was the Duke Liver dataset (\cite{macdonald2023duke}) that contained 2146 multi-parametric MRI series from 105 patients. In contrast to the AMOS22 dataset, it offered comprehensive details on data acquisition and patient demographics. Given that it exclusively contained annotations of the liver, the second external validation study specifically focused on evaluating the performance of MRISegmentator for liver segmentation. As shown in Fig.~\ref{Fig:inclusion_exclusion_dukeLiver} in the supplementary materials, we excluded MRI volumes that were not T1w MRI series and those volumes that lacked liver segmentation labels, which narrowed the dataset down to 172 T1w MRI volumes from 95 patients. Detailed information on data acquisition and patient demographics information of the Duke Liver dataset used in our study can be found in Table~\ref{tab-duke-metaInfo} in the supplementary materials. For further information about these two datasets, we refer interested readers to the respective publications (\cite{macdonald2023duke,ji2022amos}).

\subsection{Annotation}
\label{sec:anno}

The internal dataset was annotated to provide segmentation maps of 62 annotated abdominal organs and structures, thereby offering comprehensive coverage across various anatomical regions within the abdomen. These structures were organised into four different categories based on their anatomical locations and physiology: (1) Group 1 (G1) includes 15 major organs, (2) Group 2 (G2) consists of 7 blood vessels, (3) Group 3 (G3) includes 8 muscles, and (4) Group 4 (G4) contains 32 skeletal structures. For a detailed enumeration of the organs and structures within each group, the reader is referred to Section~\ref{organList} in the supplementary materials.


\begin{figure}[!htb]
    \centering
    \begin{minipage}{1\textwidth}
        \centering
        \includegraphics[width=1\linewidth]{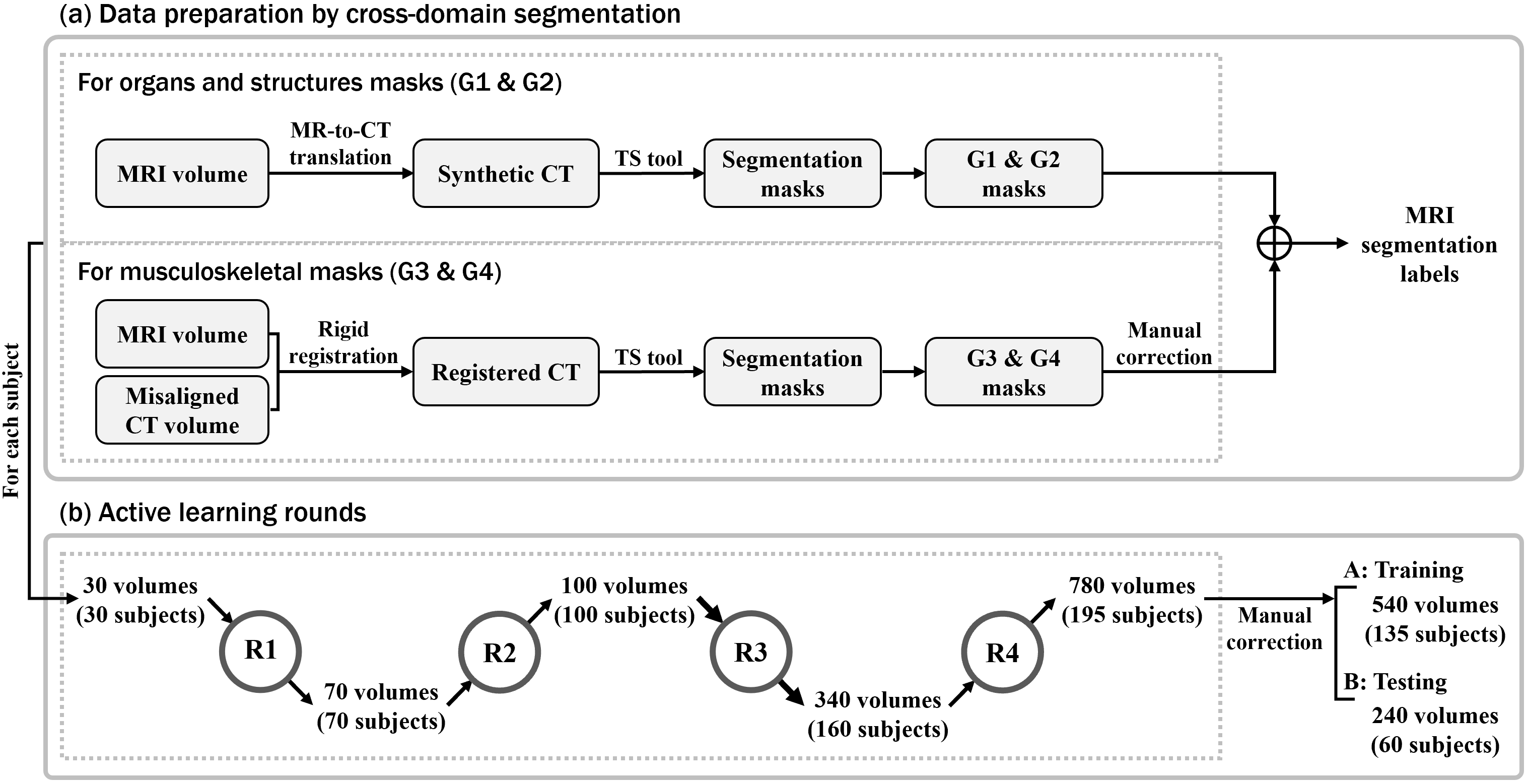} 
    \end{minipage}
\caption{The semi-automated annotation pipeline consists of two steps: (a) a data preparation step; (b) an iterative learning pipeline. First, the cross-domain segmentation method was leveraged to derive large-scale pseudo-labels. This involved synthetic translation of MRI volumes into CT, pseudo-label generation using TotalSegmentator (TS) for the synthetic CT volumes, and subsequent mapping of pseudo-labels back to the MRI volumes. Then, the iterative learning strategy refined the pseudo-annotations in each round. The upper branch of (a) was used to obtain masks for main abdominal organs and structures in G1 and G2. The bottom branch of (a) was utilized to obtain segmentation masks for musculoskeletal structures in G3 and G4. The iterative learning pipeline included multiple iterations of ``training-correction-retraining''. In each iteration, manual refinements of pseudo-labels were incorporated into the training phase to yield a powerful model, which was subsequently used to generate pseudo-labels on additional MRI volumes in the next iteration. At the end of the iterative learning pipeline, the dataset was randomly split into a training set and a testing set.}
\label{Fig:initial_segmentation} 
\end{figure}

Annotating 62 organs and structures across 780 volumes (equating to 69,248 2D slices) is a task of considerable complexity and an exceedingly labor-intensive endeavor~(\cite{park2020annotated}). As shown in Fig.~\ref{Fig:initial_segmentation}, a semi-automated annotation approach was adopted to reduce the time required for annotation. First, the ``cross-domain segmentation'' step seen in Fig.~\ref{Fig:initial_segmentation}(a) produced preliminary segmentations (also called pseudo-labels). Subsequently, these pseudo-labels were incrementally refined through an iterative learning approach over multiple iterations as shown in Fig.~\ref{Fig:initial_segmentation}(b).

The upper branch of Fig.~\ref{Fig:initial_segmentation}(a) illustrates the process of generating non-musculoskeletal masks for organs and vessels in groups G1 and G2 ($e.g.$, liver and aorta) through synthetic segmentation~(\cite{zhuang2024segmentation}). This process involved conversion of the MRI volumes into synthetic CT volumes. Subsequently, the CT-based TotalSegmentator (TS) tool (\cite{wasserthal2023totalsegmentator}) predicted the corresponding segmentation masks for these synthetic CT volumes. The masks were then mapped back to the MRI volume as pseudo-labels. The bottom branch of Fig.~\ref{Fig:initial_segmentation}(a) depicts the creation of pseudo-masks for muscle and bone structures in G3 and G4 (e.~g., spines and gluteus muscles) through the segmentation of paired CT scans. We registered the paired CT and MRI volumes together for each patient, and applied TS to the registered CT volumes to generate pseudo-labels. These pseudo-labels were manually refined by an expert grader (YZ) to ensure they accurately aligned with the underlying anatomy. Following this, we combined all pseudo-masks to form the initial segmentation masks, setting the stage for the next round of active learning.

In the second setup, the iterative learning pipeline involved several iterations of ``training-refinement-retraining'' procedure to incorporate experts' corrections and to progressively improve the segmentation masks within each round from Round 1 (R1) to Round 4 (R4). For each training round, a grader (YZ) initially refined the initial segmentations manually. Following this, an nnU-Net segmentation network (\cite{isensee2021nnu}) was trained to generate predictions for additional volumes that were utilized in the subsequent rounds. At the end of R4, comprehensive voxel-wise segmentations for each of the 62 structures were obtained for all 780 MRI volumes across 195 patients. 

The dataset was then randomly split into training (135 patients, 540 MRI volumes) and testing sets (60 patients, 240 MRI volumes). Within the training set, a radiology resident (BK) verified the annotations for one sequence per patient for a subset of 60 out of the 135 patients. Following this, the grader (YZ) manually corrected the annotations, and also reviewed and amended the annotations for the rest. In the testing set, both a senior board-certified radiologist with 30+ years of experience (RMS) and a radiology resident (BK) manually reviewed two sequences per patient. The grader (YZ) then manually refined the annotations for all testing set volumes based on their collective feedback.

\subsection{Deep Learning Model}

MRISegmentator utilized the self-configuring 3D nnUNet segmentation framework (\cite{isensee2021nnu}), which currently stands as the \textit{de-facto} standard in medical image segmentation. The nnUNet model has won multiple challenges with benchmark datasets and been robustly validated in contrast to recent novel CNN-based, transformer-based, and Mamba-based architectures (\cite{isensee2024nnu}). Built on the PyTorch framework, it automatically identified the optimal hyper-parameters for training a segmentation model, enabling it to efficiently segment target structures of interest. Models trained using this framework can be easily shared with the research community for reproducibility. During the training phase, MRISegmentator processed MRI volumes along with the corresponding segmentation masks for 62 distinct structures as input. A 3D full-resolution nnUNet was trained with five-fold cross-validation, and it resulted in five models that were used in an ensemble for prediction on the test set. The number of training epochs was set to 2000 to account for the large number of structures. The loss function was an equally weighted combination of binary cross-entropy and soft Dice losses. Optimization of this loss function was achieved using the Adam optimizer with an initial learning rate of $10^{-2}$, a polynomial learning rate scheduler, and a batch size of 2. All experiments were conducted using an NVIDIA A100 80GB GPU on the NIH Biowulf high-performance computing cluster.

\section{Experiments and Statistical Analysis}

\medskip
\noindent
\textbf{Experiments.} The performance of MRISegmentator was assessed through validation on both internal and external test datasets. For the internal test set, segmentation results produced by the proposed tool for all 62 organs and structures were analyzed. External validation of MRISegmentator was conducted on the AMOS22 MRI data subset (60 patients, 60 volumes) and the Duke Liver dataset (95 patients, 172 volumes). Additionally, MRISegmentator's performance was compared against other state-of-the-art methods on the validation set of the AMOS22 MRI subset (20 patients, 20 volumes), as reported by Ji et al.~(\cite{ji2022amos}). The segmentation accuracy was quantitatively measured using the Dice Similarity Coefficient (DSC) and Normalized Surface Distance (NSD).

An additional experiment was also conducted wherein a 3D full-resolution nnUNet model ``nnUNet-AMOS'' was trained on the AMOS22 dataset. This model then predicted segmentation masks for 13 structures in the internal dataset. As AMOS22 is currently the largest publicly available MRI dataset with labels for 13 structures, the intent of this experiment was to determine the performance of a model trained on AMOS22 and tested on the internal test dataset. The segmentations from both ``nnUNet-AMOS'' and MRISegmentator were compared to evaluate the effectiveness of each model.

\medskip
\noindent
\textbf{Statistical Analysis.} For each patient, the DSC and NSD scores were calculated for each of the 62 structures. Two statistical tests were conducted to measure the tool's performance. First, a two-sided Student's t-test was performed to compare the DSC scores between MRISegmentator and nnUNet-AMOS. A p-value $<$ .05 was considered statistically significant. Second, the performance across different MRI series (e.g., PRE vs. ART vs. VEN vs. DEL) was done to investigate the segmentation consistency across all sequences. To account for the correlations between two series from the same patient, a mixed-effects statistical model was run with the DSC scores as the dependent variable, the MRI sequence type and organ as fixed effects, incorporated a random effect for each patient. Marginal means for each type of sequence (PRE vs. ART vs. VEN vs. DEL) were obtained as well as corresponding standard errors. 
\section{Results}

\begin{figure*}
    \centering
    \begin{minipage}{0.95\textwidth}
        \centering
        \includegraphics[width=0.95\linewidth]{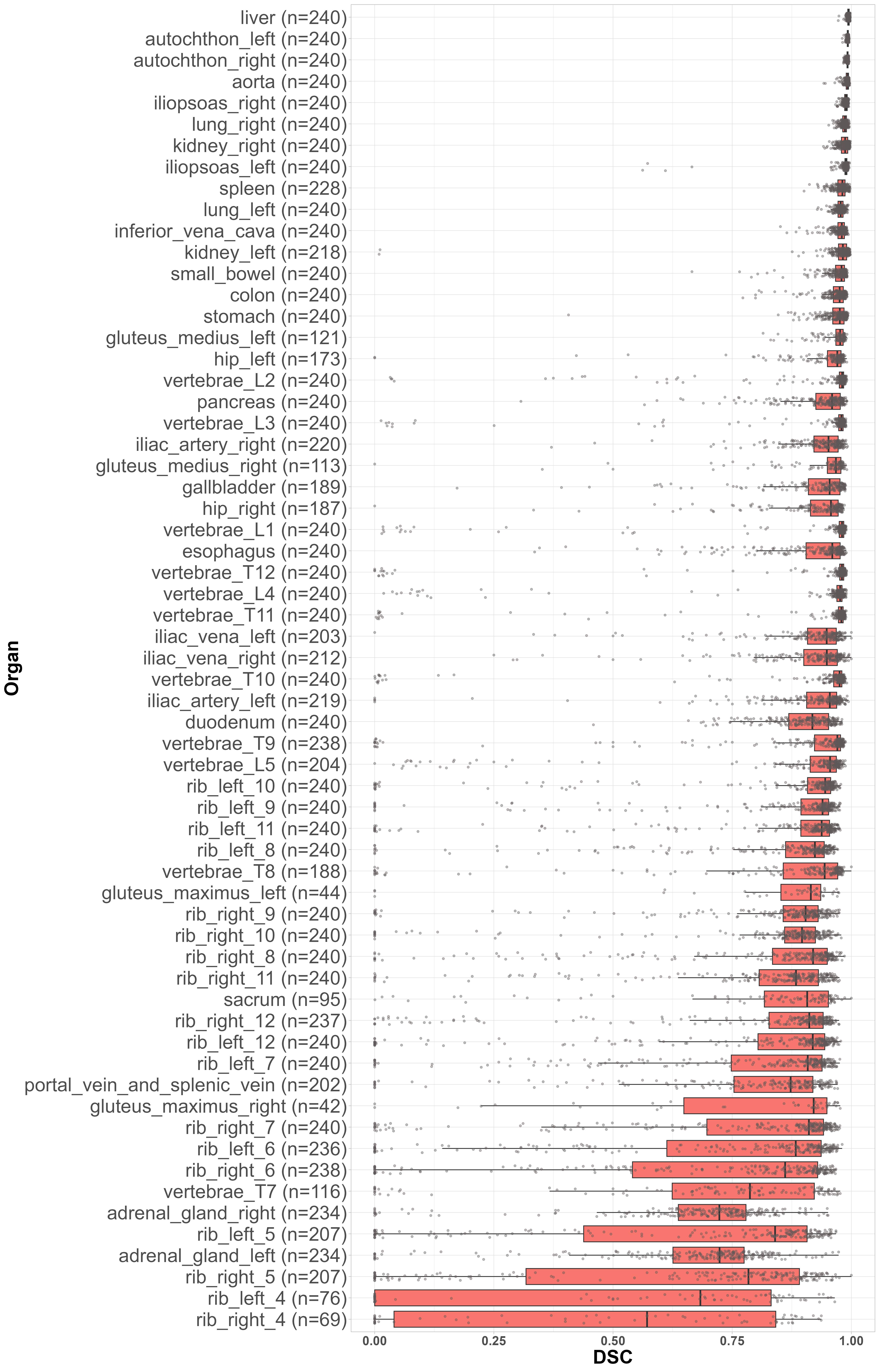} 
    \end{minipage}
\caption{Dice scores for all 62 organs and structures (ranked descending order of performance). The number of volumes in the training data subset that the structure was present in is also indicated.}
\label{Fig:dsc-segs} 
\end{figure*}

\subsection{Internal Dataset} 

\begin{figure*}[!tb]
    \begin{minipage}{1\textwidth}
        \centering
        \includegraphics[width=1\linewidth]{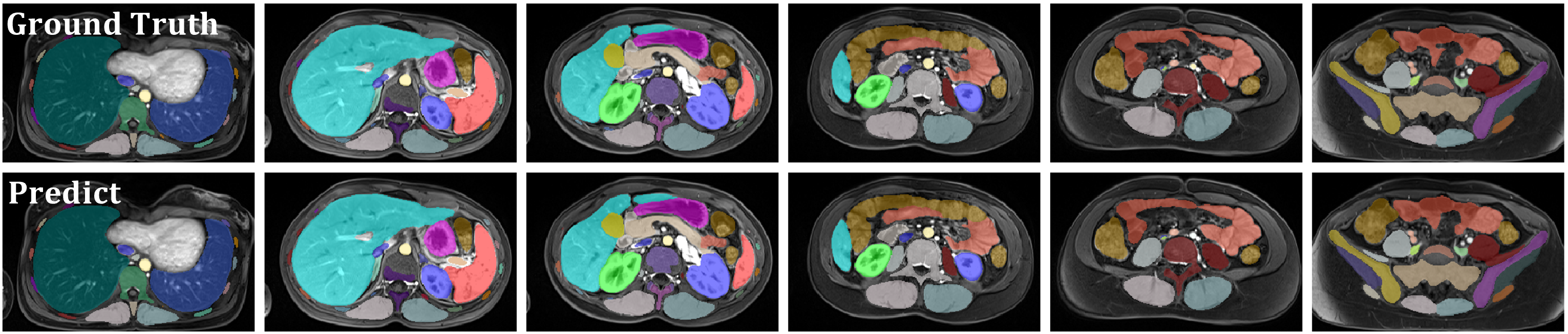}
    \end{minipage}
    \centerline{(a)}
    \begin{minipage}{1\textwidth}
        \centering
        \includegraphics[width=1\linewidth]{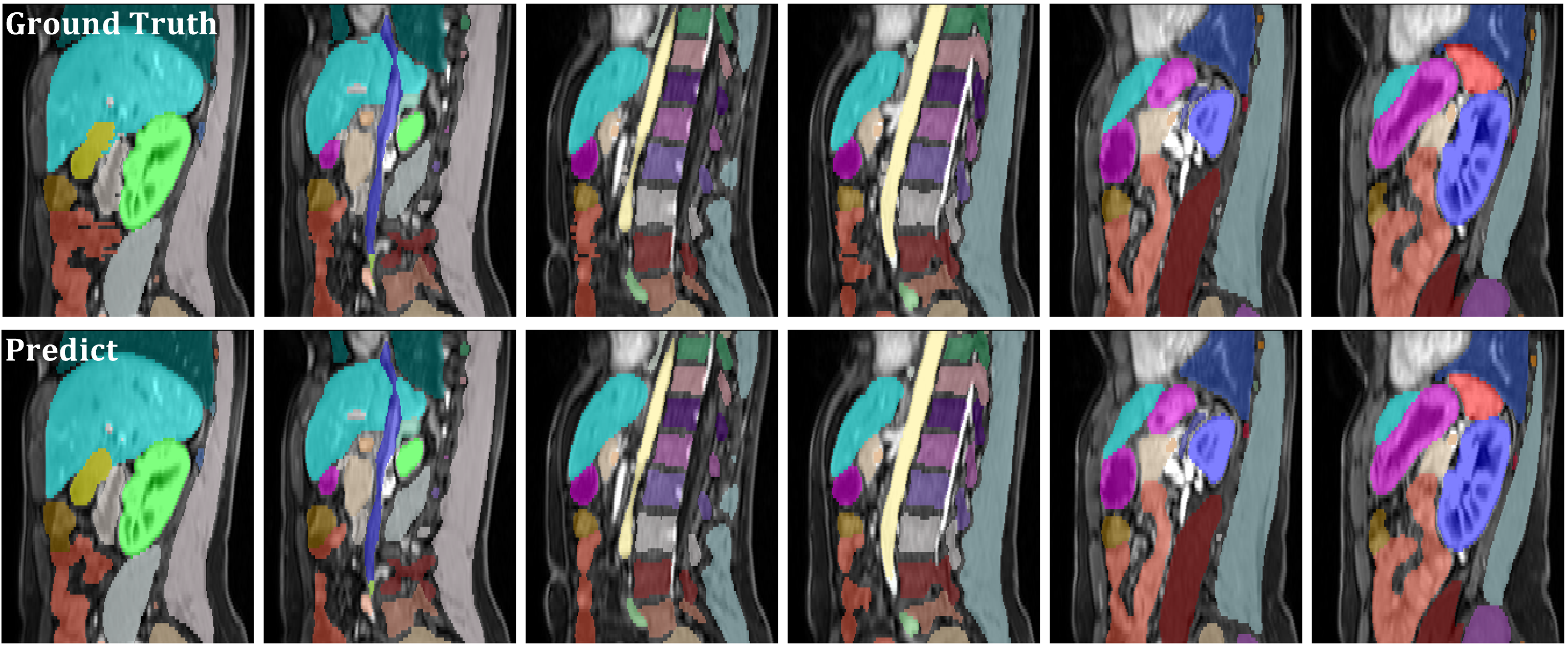}
    \end{minipage}
    \centerline{(b)}
    \begin{minipage}{1\textwidth}
        \centering
        \includegraphics[width=1\linewidth]{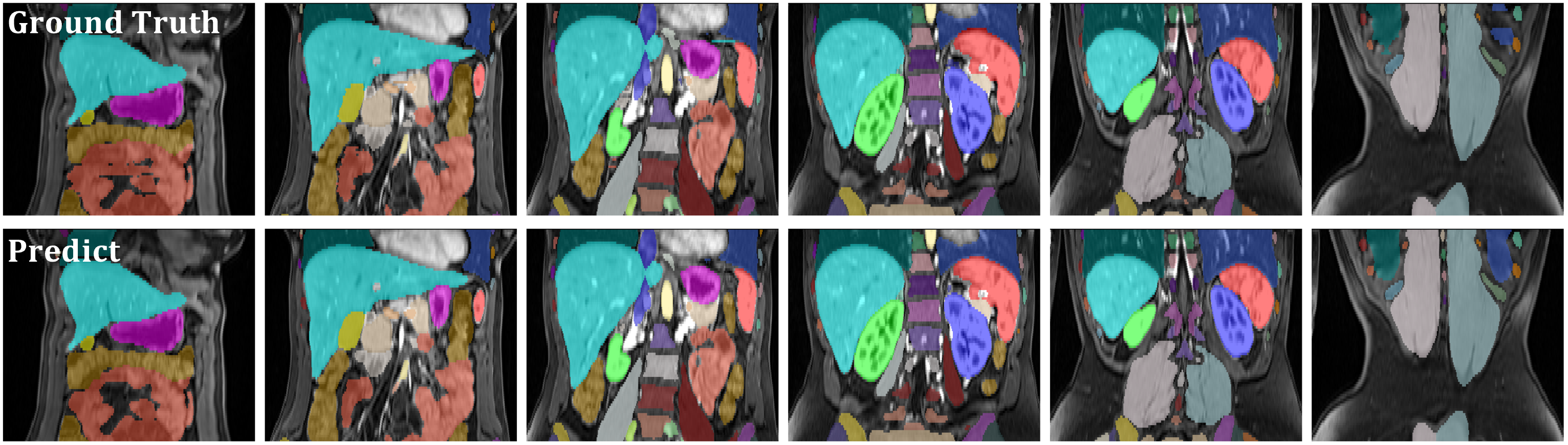}
    \end{minipage}  
    \centerline{(c)}
\caption{Ground truth and predicted segmentation results for one patient in axial, sagittal, and coronal views: (a) Axial view - Segmentation results from superior to inferior; (b) Sagittal view - Segmentation results from right to left; (c) Coronal view - Segmentation results from anterior to posterior. The color map of segmentation for each organ is provided in Fig.~\ref{Fig:organLegend} in the supplementary materials.}
\label{Fig:segment_results_3_views}
\end{figure*}

Fig.~\ref{Fig:dsc-segs} shows the mean DSC for the 62 organs and structures across all patients. Tables~\ref{tab:DSC_score} and \ref{tab:NSD_score} in the supplementary materials show the DSC and NSD scores (mean$\pm$std). Fig.~\ref{Fig:segment_results_3_views} depicts the segmentation results for a patient in the axial, sagittal, and coronal views. The color map of the segmentation mask for each organ is shown Fig.~\ref{Fig:organLegend} in the supplementary materials. 

Overall, MRISegmentator achieved an average DSC of 0.861$\pm$0.170 and a NSD of 0.924$\pm$0.163 across all 62 organs and structures in the abdomen for all scans in the internal test set. The segmentation results were also analyzed for each group of structures. The DSC and NSD were for G1 (15 major organs) 0.918$\pm$0.069 and 0.959 $\pm$ 0.067; G2 (vessels) 0.913 $\pm$ 0.105 and 0.972 $\pm$ 0.080; G3 (muscles) 0.929 $\pm$ 0.093 and 0.971 $\pm$ 0.084; and G4 (bones) 0.806 $\pm$ 0.250 and 0.886 $\pm$ 0.245, respectively.

Larger organs and structures with relatively non-complex anatomical shapes, such as the liver, spleen, kidneys and autochthonous (back) muscles in G1 and G3, achieved the highest DSC scores. Conversely, as anatomical complexity increased, structures like the small bowel, colon, stomach, and pancreas yielded lower DSC scores. The DSC scores were lowest when delineating smaller organs and structures, such as adrenal glands and ribs. There are inherent difficulties in segmenting small objects, and they are further compounded by potential patient movement and imaging artifacts. MRISegmentator encountered challenges while delineating structures located at the transition from the chest to the abdomen, or from the abdomen to the pelvis. For example in Fig.~\ref{Fig:dsc-segs}, the lowest DSC scores were seen for the thoracic ribs (rib\_4 and rib\_5) and thoracic verterbrae (T7), which are located at the transitional intersection of the lower chest and upper abdomen.

The DSC scores for the pre-contrast (PRE), arterial (ART), venous (VEN), and delayed (DEL) phases were 0.863$\pm$0.231, 0.874$\pm$0.217, 0.880$\pm$0.212, and 0.876$\pm$0.220, respectively. NSD scores were 0.922$\pm$0.208, 0.932$\pm$0.196, 0.937$\pm$0.191, 0.933$\pm$0.197, respectively. The pre-contrast T1 MRI (PRE) sequence exhibited the lowest performance, primarily due to being a pre-contrast sequence where some organs (pancreas) and structures (vessels) do not show any enhancement patterns, which affects the precise segmentation of these structures. The mixed-effects model analysis found that the differences between PRE and the ART/VEN/DEL sequences were statistically significant (\textit{p} $<$ 0.05), indicating a higher segmentation accuracy in post-contrast phases. However, no significant difference were observed among the ART/VEN/DEL sequences, which suggested a comparable segmentation performance for these three sequences. 

\begin{figure}[!t]
    \centering
    \begin{minipage}{1\textwidth}
        \centering        \includegraphics[width=1\linewidth]{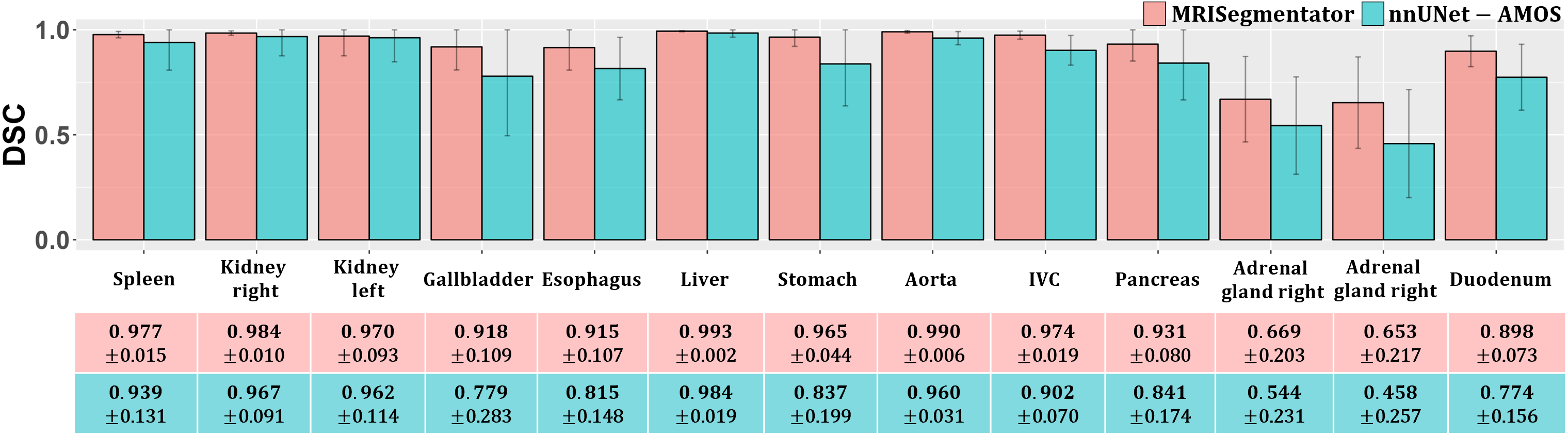} 
    \end{minipage}
\caption{Comparison of segmentation performance between MRISegmentator and nnUNet-AMOS for 13 key organs in the internal dataset. These organs were chosen as nnUNet-AMOS was trained on the AMOS22 dataset that contained 13 organ labels.}
\label{Fig:inhouse-compre} 
\end{figure}

\begin{figure*}[!htb]
    \begin{minipage}{1\textwidth}
        \centering        
        \includegraphics[width=1\linewidth]{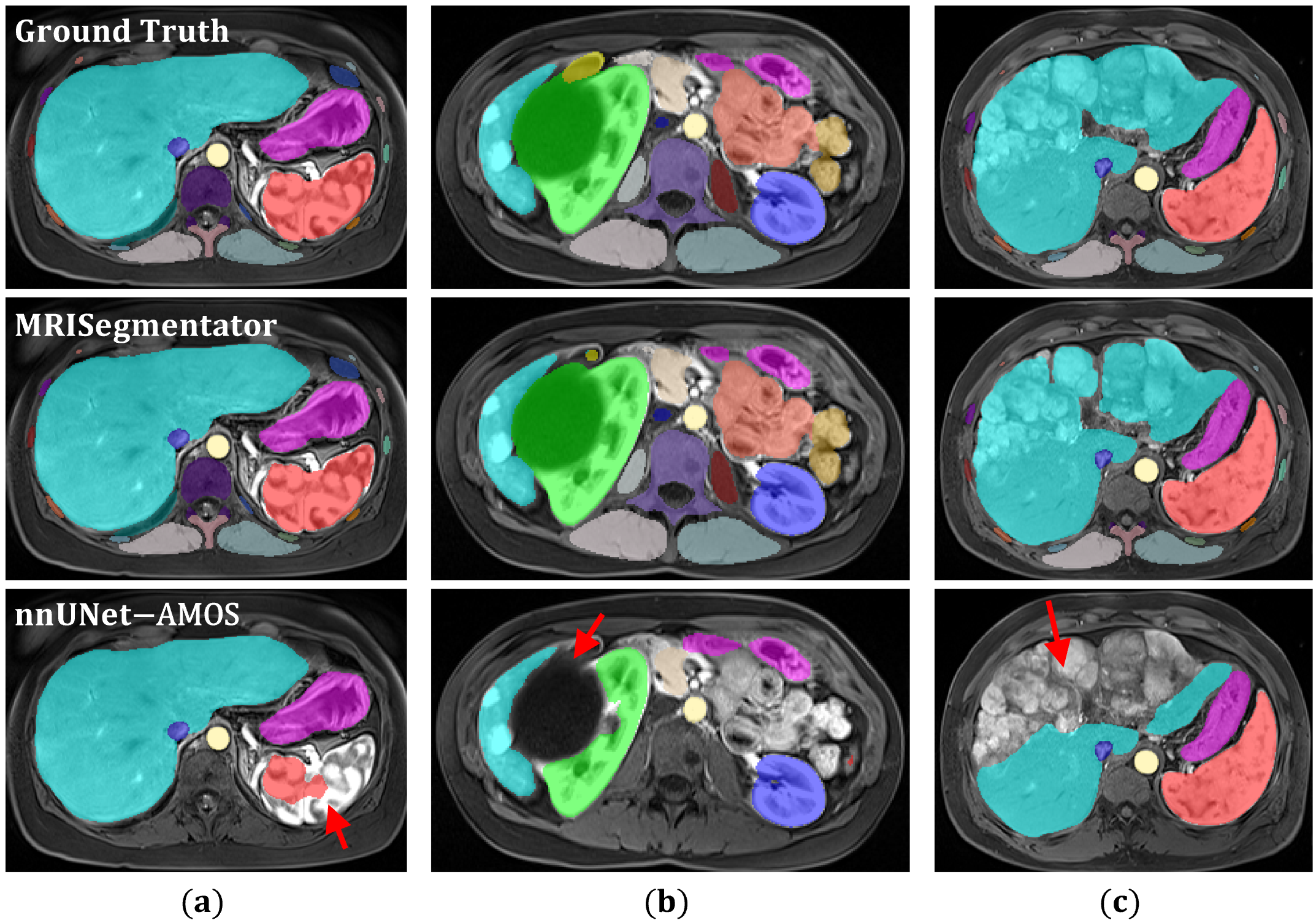}
    \end{minipage}
\caption{Comparison of MRISegmentator and nnUNet-AMOS with ground truth on segmenting structures with different types of pathology. nnUNet-AMOS struggled to segment the same regions as highlighted by the red arrows in the last row. (a) shows a patient with heterogeneous splenic attenuation during the arterial phase of enhancement; (b) shows a patient with a hypo-intense kidney lesion; (c) shows multiple liver lesions in a patient with metastatic disease.}


\label{Fig:pathCASE} 
\end{figure*}

Fig.~\ref{Fig:inhouse-compre} shows the DSC scores obtained by MRISegmentator and nnUnet-AMOS for 13 major abdominal organs in the internal dataset. Detailed NSD scores from both MRISegmentator and nnUnet-AMOS for the 13 structures are provided in Table~\ref{tab:NSD_mriseg_amos22} in the supplementary materials. Again, MRISegmentator outperformed nnUnet-AMOS by a large margin for a majority of organs, such as the gallbladder, stomach, and duodenum. For large organs, such as the kidneys and the liver, the performance differences were subtle. For DSC scores, MRISegmentator significantly outperformed nnUnet-AMOS (\textit{p} $<$ .0038, two-sided t-test with Bonferroni correction). One contributing factor to the decreased segmentation performance of nnUnet-AMOS was the presence of a variety of pathological cases in the internal dataset, including liver tumors and pancreatic cysts. The AMOS22 dataset does not contain training data from patients with different pathologies. Fig.~\ref{Fig:pathCASE} presents several instances where nnUNet-AMOS failed to accurately segment regions affected by pathological changes, leading to a decrease in DSC scores in these cases.

\subsection{External Dataset}

\begin{figure}[!t]
    \centering
    \begin{minipage}{1\textwidth}
        \centering    \includegraphics[width=1\linewidth]{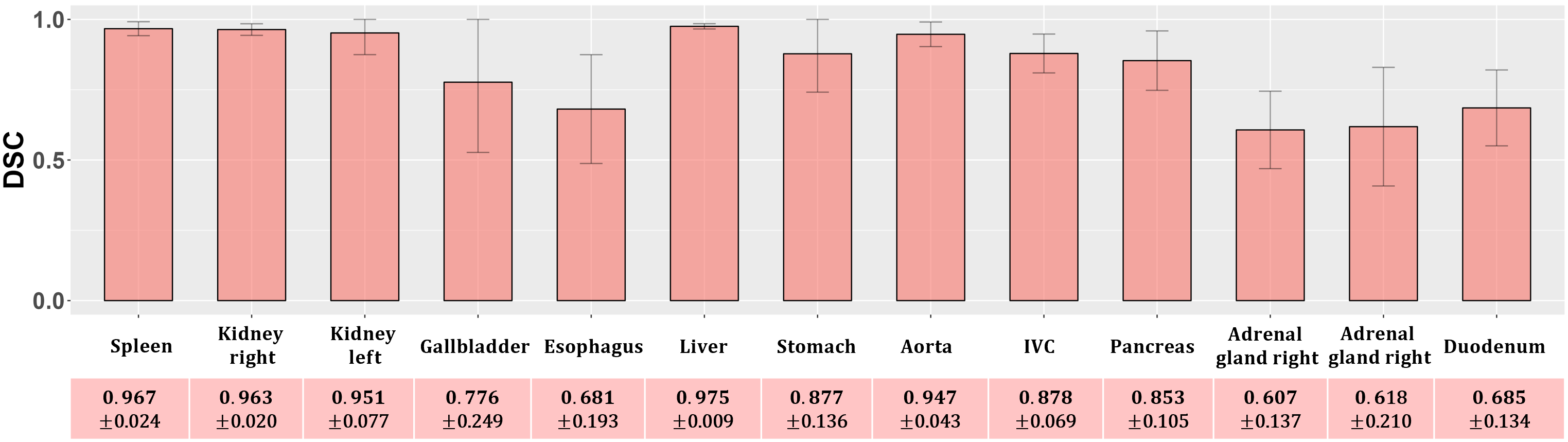} 
    \end{minipage}
\caption{The segmentation performance of MRISegmentator for 13 major organs in the train + validation subsets of the AMOS22 dataset.}
\label{Fig:dsc_ams_on_amos22} 
\end{figure}

Fig.~\ref{Fig:dsc_ams_on_amos22} shows the segmentation performance of MRISegmentator for each of the 13 organs across all 60 patients in the AMOS22 dataset. Overall, MRISegmentator achieved an average DSC of 0.829$\pm$0.133 across 13 organs. Table~\ref{tab:NSD_mriseg_amos22} in the supplementary materials shows the DSC and NSD scores for each individual organ. On the validation set of the AMOS22 MRI subset (20 patients, 20 volumes), MRISegmentator obtained a mean DSC of 0.844, which the second-best performance compared to other methods that were directly trained on the AMOS22 dataset (CoTr~(\cite{xie2021cotr}): 0.775, nnFormer~(\cite{zhou2021nnformer}): 0.806, Swin-UNetr~(\cite{hatamizadeh2021swin}): 0.757, UNet~(\cite{ronneberger2015u}): 0.856, UNetr~(\cite{hatamizadeh2022unetr}): 0.753, VNet~(\cite{milletari2016v}: 0.837)), as reported by Ji et al.~(\cite{ji2022amos}). This observation is remarkable as MRISegmentator was not directly trained with the AMOS22 dataset.

On the Duke Liver dataset, the DSC scores for the PRE, ART, and VEN phases were 0.912$\pm$0.055, 0.946$\pm$0.012, 0.942$\pm$0.019, respectively. The NSD scores were 0.903$\pm$0.070, 0.929$\pm$0.024, and 0.956$\pm$0.017, respectively. MRISegmentator achieved DSC scores of $>$ 0.9 across all three sequences (PRE, ART, and VEN). The lowest DSC score was observed in the PRE squence, attributable to the reduced contrast inherent in pre-contrast imaging.

\section{Discussion}

The proposed MRISegmentator is a multi-organ and structure segmentation tool that can automatically segment 62 abdominal organs and structures. MRISegmentator was extensively validated on an internal dataset and two external datasets from different institutions. Empirical results demonstrated that the model achieved average DSC scores of exceeding 0.90 for 15 major organs (except the adrenals glands), 7 blood vessels, and 8 muscles, and 0.80 for 35 bones. Furthermore, external validations showed its robust performance across two large abdominal MRI datasets from different institutions. MRISegmentator obtained an average DSC of 0.829$\pm$0.133 for 13 organs in the AMOS22 dataset, and attained DSC scores of $>$ 0.9 across all three sequences (PRE, ART, and VEN) in the external Duke Liver dataset.

The segmentation performance on the external AMOS22 dataset was not as high as the results observed in the internal dataset. MRISegmentator achieved a DSC score of $\ge$ 0.85 for the majority of organs, with the exception of several smaller organs such as the gallbladder, esophagus, adrenal glands, and duodenum. This discrepancy can be attributed to the non-T1w and non-axial volumes in the AMOS22 dataset\footnote{Due to the absence of acquisition information in the AMOS22 dataset, excluding non-T1w images was challenging.}. These series types were unseen during the training phase of MRISegmentator and diverged from the training data distribution. Nevertheless, MRISegmentator showed its robustness to different coronal volumes. Fig.~\ref{Fig:amos_coronal_bad_example} in the supplementary materials demonstrates the segmentation results of a patient in coronal-view T1w MRI from the AMOS22 dataset. On the validation set of the AMOS22 MRI subset (20 patients, 20 volumes), though MRISegmentator was not directly trained with the AMOS22 dataset, it still came in second when compared against other state-of-the-art methods that were trained directly on AMOS22 training set. This showcases the robust segmentation capabilities of MRISegmentator on a dataset obtained from an external institution.

Furthermore, the segmentation performance was sub-optimal for ribs and spinal vertebrae, particularly in the upper and lower abdomen near transitional zones of the chest and pelvis. Fig.~\ref{Fig:amos_coronal_bad_example} in the supplementary materials illustrates this issue. Results from the coronal volumes in the AMOS22 dataset contained a portion of the lower chest, and showed that parts of the lung, ribs, and spinal vertebrae were erroneously segmented. These errors occurred as the chest MRIs were not included in the training data. Chest MRI studies are infrequently acquired in our institution as patient motion from breathing can corrupt the MRI sequences.

To the best of our knowledge, multi-structure segmentation on the scale of our work has only been explored in a few recent studies. Chen $et$ $al.$~(\cite{chen2020fully}) proposed a 2D UNet-based method to segment 10 major abdominal organs and bones using their internal dataset of 102 patients, but this study covered only a limited number of structures. Ji $et$ $al.$~(\cite{ji2022amos}) provided the first publicly available MRI dataset containing multi-structure labels as part of the AMOS22 challenge, and it featured annotations for 13 structures across 60 MRI volumes. However, the limited number of patients in this dataset posed significant challenges for model generalization, and lacked data acquisition parameters and patient demographic information. The Duke Liver dataset~(\cite{macdonald2023duke}), while providing detailed sequence information, was primarily released for automated classification of MRI sequences and exclusively contains only liver segmentations. More recently, several concurrent works on multi-organ and structure segmentation for MRI have emerged. Gu $et$ $al.$~(\cite{gu2024segmentanybone}) proposed a universal bone segmentation network to segment skeletal structures in MRI. In parallel, Zhou $et$ $al.$~(\cite{zhou2024mrannotator}) proposed MRAnnotator to segment 49 structures in the whole body. Gei{\ss}ler $et$ $al.$~(\cite{Kai2023GIN}) proposed a TS-like tool for T1 Dixon MRI using global intensity non-linear data augmentation. However, these tools encompassed fewer structures in contrast to our work. In addition, the datasets or models were not publicly available. In our study, we have created a robust tool to segment 62 organs and structures, and it has been comprehensively evaluated.


There are some limitations to our study. First, the internal dataset included patients with a broad range of pathologies, and an analysis of this effect posed on the segmentation accuracy of MRISegmentator has not been evaluated. Second, this study solely focused on pre-contrast and dynamic contrast enhanced T1-weighted MRI series. Other MRI sequences, such as T1 Dixon MRI, T1 in-phase/opposed phase, T2-weighted and T2 Fat-Suppressed, were not considered and are the subject of future work. Additionally, the current work only focused on axially acquired series, but as described above, it was robust to coronal and sagittal sequences as well. Third, chest or pelvic MRI studies were not considered. The former is less frequently acquired and the latter is a potential avenue for future work~(\cite{zhuang2024segmentation}). Fourth, we did not compare our 3D nnUNet model against other approaches, such as transformer-based segmentation models or foundation models. The nnUNet framework was robustly validated on a variety of segmentation challenges and it is the \textit{de-facto} standard for medical segmentation tasks (\cite{isensee2024nnu}). Finally, the testing set was reviewed by a single radiologist and inter-rater variability of segmentations was not performed.

In conclusion, a multi-parametric T1-weighted MRI segmentation tool called MRISegmentator that accurately and reliably segments 62 organs and structures in body MRI has been presented. The tool has the potential to be easily integrated into the clinical workflow for various tasks, such as abnormality detection, early detection of cancer, and opportunistic screening with body composition measurements among others. MRISegmentator is efficient and accurate, and it can save time when initial organ or structure segmentations are needed for downstream tasks, such as tracking interval changes in size.

\bmhead{Acknowledgments}

This work was supported by the Intramural Research Program of the National Institutes of Health (NIH) Clinical Center (project number 1Z01 CL040004). This work used the computational resources of the NIH HPC Biowulf cluster.

\bmhead{Declarations}

The authors declare the following financial interests/personal relationships which may be considered as potential competing interests: Ronald M. Summers reports a relationship with Ping An (CRADA) that includes: funding grants. Co-author RMS receives royalties from iCAD, Philips, ScanMed, PingAn, MGB, and Translation Holdings.

\bigskip

%
\bibliography{sn-bibliography}

\clearpage

\section{Supplemental Materials}

\subsection{The internal dataset}

\begin{figure}[H]
    \centering
    \begin{minipage}{1\textwidth}
        \centering
        \includegraphics[width=0.75\linewidth]{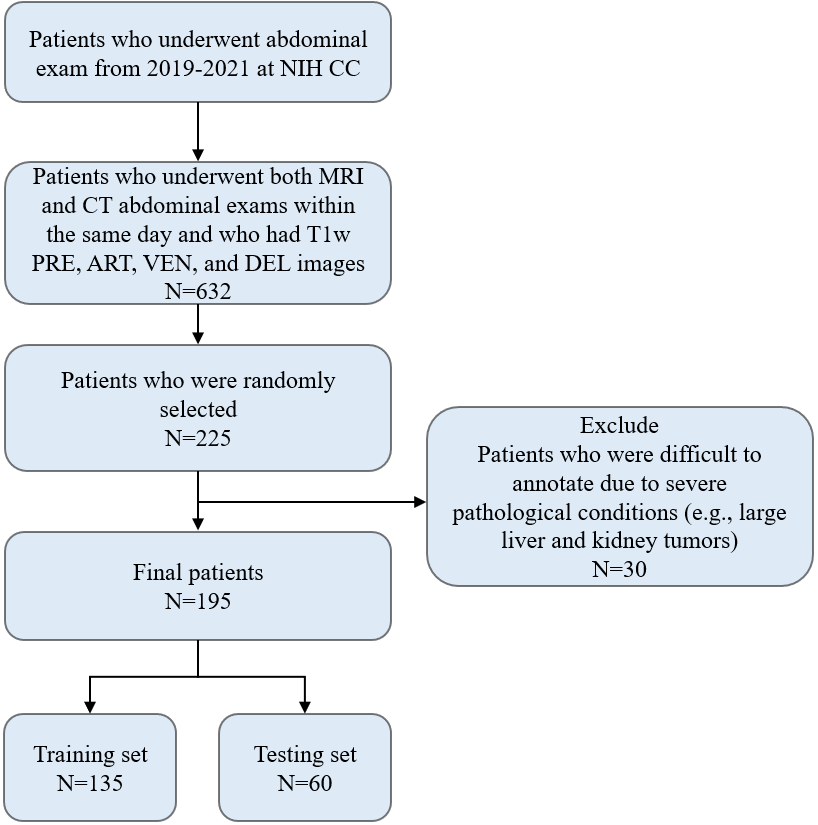} 
    \end{minipage}
\caption{The Standards for Reporting of Diagnostic Accuracy (STARD) chart that details the inclusion and exclusion criteria for the internal dataset.}
\label{Fig:inclusion_exclusion_internal} 
\end{figure}

Fig.~\ref{Fig:inclusion_exclusion_internal} shows the STARD chart detailing the inclusion and exclusion criteria for the internal dataset. Table~\ref{tab-internal-metaInfo} shows a summary of the acquisition information for the internal dataset. 

\begin{table*}[!t]
  \centering
  \caption{Details on the MRI acquisition and patient demographics for the internal T1-weighted MRI dataset.}
    \resizebox{0.8\textwidth}{!}{%
    \begin{tabular}{|l|c|c|}
    \toprule
    \multicolumn{3}{|c|}{\textbf{Internal Dataset}} \\
    \midrule
    \multicolumn{3}{|c|}{\textbf{Acquisition information (T1w PRE, ART, VEN, DEL)}} \\
    \midrule
          & Train & Test \\
    \midrule
    Number of volumes (patients)    &540 (135)      &240 (60)  \\
    \midrule
    Width range         &228-320       &320-320  \\
    \midrule
    Height range         &240-320       &220-290  \\
    \midrule
    Number of slices    &80-96       &80-104  \\
    \midrule
    Slice thickness range (mm)  &3.0-3.3       &3  \\
    \midrule
    Pixel spacing range (mm)    &(0.937-1.468)x(0.937-1.468)    &(1.0-1.375)x(1.0-1.375)  \\
    \midrule
    TR (msec)                   &3.65-3.88       &3.65-3.81  \\
    \midrule
    TE (msec)                   &1.67-1.92       &1.69-1.87  \\
    \midrule
    Flip Angle                  &10       &10  \\
    \midrule
    Field strength              &1.5, 3       &1.5, 3  \\
    \midrule
    \multicolumn{3}{|c|}{\textbf{Demographics information}} \\
    \midrule
          & Train & Test \\
    \midrule
    Age, mean+std       &54.67$\pm$16.27       &51.12$\pm$14.39  \\
    \midrule
    min, max            &16, 87       &23, 83  \\
    \midrule
    Gender ratio (female: male)      &72:63       &26:34  \\
    \midrule
    White               &100       &53  \\
    \midrule
    Black             &13       &1  \\
    \midrule
    Asian               &10       &2  \\
    \midrule
    Native Americans    &1       &0  \\
    \midrule
    Native Hawaiian or other pacific islander &1       &0  \\
    \midrule
    Multiracial         &4       &1  \\
    \midrule
    Unknown             &6       &3  \\
    \bottomrule
    \end{tabular}%
  \label{tab-internal-metaInfo}}
\end{table*}%

\subsection{The Duke Liver dataset} 

\begin{figure}[!h]
    \centering
    \begin{minipage}{1\textwidth}
        \centering
        \includegraphics[width=0.55\linewidth]{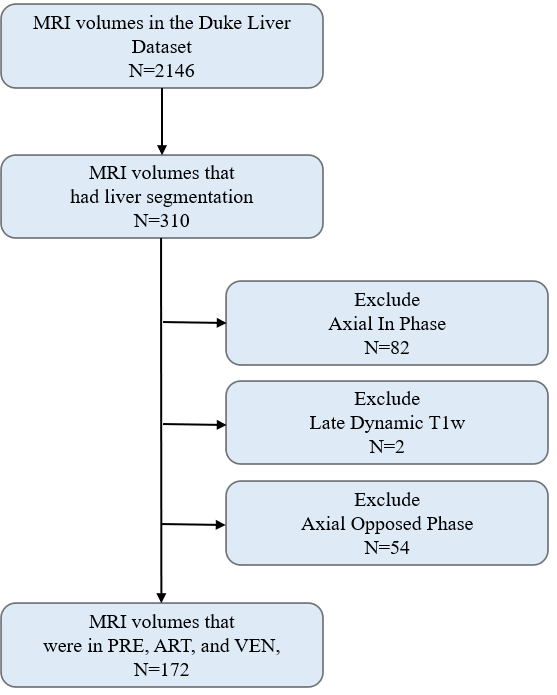} 
    \end{minipage}
\caption{The STARD chart detailing the inclusion and exclusion criteria for the external Duke Liver dataset.}
\label{Fig:inclusion_exclusion_dukeLiver} 
\end{figure}

Fig. \ref{Fig:inclusion_exclusion_dukeLiver} shows the STARD chart detailing the inclusion and exclusion criteria for the external Duke Liver dataset used in this study. Table~\ref{tab-duke-metaInfo} shows a summary of the acquisition parameters and demographics information for the external Duke Liver dataset.

\begin{table*}[!t]
  \centering
  \caption{Acquisition and demographics information for the subset of the Duke Liver Dataset used in our study}
    \begin{adjustbox}{max width=\textwidth}
    \begin{tabular}{|l|c|l|c|}
    \toprule
    \multicolumn{4}{|c|}{Duke Liver Dataset} \\
    \midrule
    \multicolumn{2}{|c|}{Acquisition information (T1w PRE, ART, VEN)} & \multicolumn{2}{c|}{Demographics information} \\
    \midrule
    Number of series (patients)		& 177 (95)    					& Age, mean$\pm$std 			& 55.66$\pm$20.68  \\
    Width range 					& 256-576       				& Age, min, max 			& 3, 79				\\
    Heigh range 					& 256-576      					& Gender ratio 				& 27:65 			\\
    Number of slices 				& 44-168      					& White 					& -					\\
    Slice thickness range (mm) 		& 3.0-7.0     					& African American 			& -					\\
    Pixel spacing range (mm) 		& (0.683-1.679)x(0.683-1.679)	& Aisan 					& - 					\\
    TR (msec) 						& 3.66-7.49       				& Native American 			& - 						\\
    TE (msec) 						& 1.07-3.12     				& Native Hawaiian or other pacific islander & - \\
    Flip Angle 						& 9-12      					& Multiracial 				& - 						\\
    Field strength 					& 1.5, 3      					& unknow 					& - 						\\
    \bottomrule
    \end{tabular}
    \end{adjustbox}
  \label{tab-duke-metaInfo}%
\end{table*}

\subsection{The acquisition information for the internal CT dataset}
Table~\ref{tab:ct_info} shows a summary of the acquisition information for the internal CT dataset used for the annotation.

\begin{table*}[]
\centering
\caption{The acquisition information for the CT data.}
\begin{adjustbox}{max width=\textwidth}
\begin{tabular}{|c|c|c|c|c|c|c|}
\hline
Scanner                    & Manufacture & Width range & Height range & Number of slices & Pixel spacing range (mm) & Slice thickness range (mm) \\ \hline
Biograph128/SOMATOM\_Force & Siemens     & 512         & 512          & 47-806           & 0.675 - 0.976       & 1.0-5.0               \\ \hline
\end{tabular}%
\end{adjustbox}
\label{tab:ct_info}
\end{table*}

\subsection{MRISegmentator - 62 abdominal organs and structures}
\label{organList}

The list below enumerates the 62 main organs and structures of interest, divided into 4 different groups based on their anatomical locations and physiology: 

\begin{itemize}
    \item \textbf{Group 1 (G1)} includes 15 main organs and structures in the abdomen: spleen, kidney\_right, kidney\_left, gallbladder, liver, esophagus, stomach,  pancreas, adrenal gland\_right, adrenal gland\_left, duodenum, lung\_right, lung\_left, small bowel, colon. 
    \item \textbf{Group 2 (G2)} includes 7 vessels: aorta, inferior vein cava, portal vein and splenic vein, iliac artery\_left, iliac artery\_right, iliac vein\_left, iliac vein\_right.
    \item \textbf{Group 3 (G3)} includes 8 muscles: gluteus maximus\_left, gluteus maximus\_right, gluteus medius\_left, gluteus medius\_right,  autochthon\_left, autochthon\_right, iliopsoas\_left, iliopsoas\_right. 
    \item  \textbf{Group 4 (G4)} includes 32 bones: hip\_left, hip\_right, sacrum, rib\_left\_4, rib\_left\_5,  rib\_left\_6, rib\_left\_7, rib\_left\_8, rib\_left\_9, rib\_left\_10, rib\_left\_11, rib\_left\_12, rib\_right\_4, rib\_right\_5, rib\_right\_6, rib\_right\_7, rib\_right\_8, rib\_right\_9, rib\_right\_10, rib\_right\_11, rib\_right\_12, vertebrae\_T7.vertebrae\_T8, vertebrae\_T9, vertebrae\_T10, vertebrae\_T11, vertebrae\_T12, vertebrae\_L1, vertebrae\_L2, vertebrae\_L3, vertebrae\_L4, vertebrae\_L5. 
\end{itemize}

\subsection{Tabular results}

Tables~\ref{tab:DSC_score} and \ref{tab:NSD_score} detail the Dice Similarity Coefficient (DSC) and Normalized Surface Distance (NSD obtained by MRISegmentator for all the 62 organs and structures in the internal dataset. 

Table~\ref{tab:NSD_mriseg_amos22} shows DSC and NSD comparison results for MRISegmentator and nnUNet-AMOS on the internal dataset, and MRISegmentator's performance on the AMOS22 dataset.

\begin{table*}[!htbp]
  \centering
  \caption{MRISegmentator - DSC of 62 organs and structures in abdomen.}
    \begin{adjustbox}{max width=\textwidth}
    \begin{tabular}{|c|l|c|c|l|c|c|l|c|c|l|c|}
    \toprule
          & \multicolumn{1}{c|}{Organ \& structure} & DSC(mean$\pm$std) &       & \multicolumn{1}{c|}{Organ \& structure} & DSC(mean$\pm$std) &       & \multicolumn{1}{c|}{Organ \& structure} & DSC(mean$\pm$std) &       & \multicolumn{1}{c|}{Organ \& structure} & DSC(mean$\pm$std) \\
    \midrule
    \multirow{16}[32]{*}{G1} & spleen    & 0.977$\pm$0.015 & \multirow{7}[14]{*}{G2} & aorta    & 0.990$\pm$0.006  & \multirow{16}[32]{*}{G4} & hip\_left    & 0.936$\pm$0.121    & \multirow{16}[32]{*}{G4} & rib\_left\_6    & 0.848$\pm$0.249 \\
\cmidrule{2-3}\cmidrule{5-6}\cmidrule{8-9}\cmidrule{11-12}          & lung\_left    			& 0.976$\pm$0.009    &       & inferior\_vein\_cava    						& 0.974$\pm$0.019    &       & hip\_right    	& 0.917$\pm$0.113    &       & rib\_left\_7    	& 0.782$\pm$0.258 \\
\cmidrule{2-3}\cmidrule{5-6}\cmidrule{8-9}\cmidrule{11-12}          & lung\_right    			& 0.984$\pm$0.006    &       & portal\_vein\_and\_splenic\_vein				& 0.777$\pm$0.239    &       & sacrum    		& 0.806$\pm$0.245    &       & rib\_left\_8    	& 0.834$\pm$0.215 \\
\cmidrule{2-3}\cmidrule{5-6}\cmidrule{8-9}\cmidrule{11-12}          & kidney\_right    			& 0.984$\pm$0.010    &       & iliac\_artery\_right    						& 0.931$\pm$0.075    &       & vertebrae\_T7	& 0.682$\pm$0.307    &       & rib\_left\_9    	& 0.846$\pm$0.234 \\
\cmidrule{2-3}\cmidrule{5-6}\cmidrule{8-9}\cmidrule{11-12}          & kidney\_left    			& 0.970$\pm$0.093    &       & iliac\_artery\_left    						& 0.904$\pm$0.153    &       & vertebrae\_T8   	& 0.833$\pm$0.270    &       & rib\_left\_10    	& 0.848$\pm$0.249 \\
\cmidrule{2-3}\cmidrule{5-6}\cmidrule{8-9}\cmidrule{11-12}          & gallbladder				& 0.918$\pm$0.109    &       & iliac\_vein\_left    						& 0.909$\pm$0.123    &       & vertebrae\_T9 	& 0.874$\pm$0.247    &       & rib\_left\_11    	& 0.836$\pm$0.253  \\
\cmidrule{2-3}\cmidrule{5-6}\cmidrule{8-9}\cmidrule{11-12}          & liver    					& 0.993$\pm$0.002    &       & iliac\_vein\_right    						& 0.908$\pm$0.118    &       & vertebrae\_T10	& 0.905$\pm$0.230    &       & rib\_left\_12    	& 0.783$\pm$0.286 \\
\cmidrule{2-6}\cmidrule{8-9}\cmidrule{11-12}          				& esophagus    				& 0.915$\pm$0.107    & \multirow{9}[18]{*}{G3} & gluteus\_medius\_right		& 0.924$\pm$0.132    &       & vertebrae\_T11	& 0.914$\pm$0.222    &       & rib\_right\_4		& 0.482$\pm$0.353 \\
\cmidrule{2-3}\cmidrule{5-6}\cmidrule{8-9}\cmidrule{11-12}          & stomach    				& 0.965$\pm$0.044    &       & gluteus\_medius\_left    					& 0.827$\pm$0.239    &       & vertebrae\_T12	& 0.915$\pm$0.229    &       & rib\_right\_5    	& 0.626$\pm$0.336 \\
\cmidrule{2-3}\cmidrule{5-6}\cmidrule{8-9}\cmidrule{11-12}          & pancreas   				& 0.931$\pm$0.080    &       & gluteus\_maximus\_left   					& 0.827$\pm$0.239    &       & vertebrae\_L1   	& 0.915$\pm$0.216   &       & rib\_right\_6  	& 0.707$\pm$0.311 \\
\cmidrule{2-3}\cmidrule{5-6}\cmidrule{8-9}\cmidrule{11-12}          & adrenal\_gland\_right   	& 0.669$\pm$0.203    &       & gluteus\_maximus\_right   					& 0.771$\pm$0.278    &       & vertebrae\_L2   	& 0.933$\pm$0.157   &       & rib\_right\_7   	& 0.762$\pm$0.286 \\
\cmidrule{2-3}\cmidrule{5-6}\cmidrule{8-9}\cmidrule{11-12}          & adrenal\_gland\_left   	& 0.653$\pm$0.217    &       & autochthon\_left   							& 0.992$\pm$0.001    &       & vertebrae\_L3   	& 0.931$\pm$0.182   &       & rib\_right\_8   	& 0.811$\pm$0.245 \\
\cmidrule{2-3}\cmidrule{5-6}\cmidrule{8-9}\cmidrule{11-12}          & duodenum    				& 0.898$\pm$0.073    &       & autochthon\_right   							& 0.991$\pm$0.002    &       & vertebrae\_L4   	& 0.915$\pm$0.213   &       & rib\_right\_9   	& 0.825$\pm$0.223 \\
\cmidrule{2-3}\cmidrule{5-6}\cmidrule{8-9}\cmidrule{11-12}          & small\_bowel   			& 0.968$\pm$0.036    &       & iliopsoas\_right   							& 0.987$\pm$0.004    &       & vertebrae\_L5   	& 0.854$\pm$0.256   &       & rib\_right\_10   	& 0.820$\pm$0.232 \\
\cmidrule{2-3}\cmidrule{5-6}\cmidrule{8-9}\cmidrule{11-12}          & colon   					& 0.965$\pm$0.032    &       & iliopsoas\_left   							& 0.981$\pm$0.049    &       & rib\_left\_4   	& 0.516$\pm$0.365   &       & rib\_right\_11   	& 0.807$\pm$0.223 \\
\cmidrule{2-3}\cmidrule{5-6}\cmidrule{8-9}\cmidrule{11-12}          & -   						& -   				 &       & -   											& -   				 &       & rib\_left\_5   	& 0.660$\pm$0.337   &       & rib\_right\_12   	& 0.788$\pm$0.277 \\
    \bottomrule
    \end{tabular}%
    \end{adjustbox}
  \label{tab:DSC_score}%
\end{table*}%

\begin{table*}[!htbp]
  \centering
  \caption{NSD results of the proposed MRISegmentator tool for 62 organs and structures in the internal dataset.}
    \begin{adjustbox}{max width=\textwidth}
    \begin{tabular}{|c|l|c|c|l|c|c|l|c|c|l|c|}
    \toprule
          & \multicolumn{1}{c|}{Organ \& structure} & NSD(mean$\pm$std) &       & \multicolumn{1}{c|}{Organ \& structure} & NSD(mean$\pm$std) &       & \multicolumn{1}{c|}{Organ \& structure} & NSD(mean$\pm$std) &       & \multicolumn{1}{c|}{Organ \& structure} & NSD(mean$\pm$std) \\
    \midrule
    \multirow{16}[32]{*}{G1} 										& spleen    				& 0.982$\pm$0.019 & \multirow{7}[14]{*}{G2}  & aorta    					& 0.999$\pm$0.003  & \multirow{16}[32]{*}{G4} & hip\_left    & 0.980$\pm$0.112    & \multirow{16}[32]{*}{G4} & rib\_left\_6    & 0.824$\pm$0.298 \\
\cmidrule{2-3}\cmidrule{5-6}\cmidrule{8-9}\cmidrule{11-12}          & lung\_left    			& 0.974$\pm$0.017    &       & inferior\_vein\_cava    						& 0.996$\pm$0.014    &       & hip\_right    	& 0.987$\pm$0.075    &       & rib\_left\_7    	& 0.870$\pm$0.247 \\
\cmidrule{2-3}\cmidrule{5-6}\cmidrule{8-9}\cmidrule{11-12}          & lung\_right    			& 0.990$\pm$0.010    &       & portal\_vein\_and\_splenic\_vein				& 0.879$\pm$0.228   &        & sacrum    		& 0.912$\pm$0.224    &       & rib\_left\_8    	& 0.910$\pm$0.215  \\
\cmidrule{2-3}\cmidrule{5-6}\cmidrule{8-9}\cmidrule{11-12}          & kidney\_right    			& 0.988$\pm$0.015    &       & iliac\_artery\_right    						& 0.989$\pm$0.033    &       & vertebrae\_T7    	& 0.841$\pm$0.317    &       & rib\_left\_9    	& 0.913$\pm$0.232 \\
\cmidrule{2-3}\cmidrule{5-6}\cmidrule{8-9}\cmidrule{11-12}          & kidney\_left    			& 0.968$\pm$0.091    &       & iliac\_artery\_left    						& 0.967$\pm$0.142    &       & vertebrae\_T8    	& 0.907$\pm$0.244    &       & rib\_left\_10    	& 0.910$\pm$0.248 \\
\cmidrule{2-3}\cmidrule{5-6}\cmidrule{8-9}\cmidrule{11-12}          & gallbladder				& 0.966$\pm$0.069    &       & iliac\_vein\_left    						& 0.987$\pm$0.081    &       & vertebrae\_T9    	& 0.928$\pm$0.225    &       & rib\_left\_11    	& 0.903$\pm$0.251 \\
\cmidrule{2-3}\cmidrule{5-6}\cmidrule{8-9}\cmidrule{11-12}          & liver    					& 0.993$\pm$0.009    &       & iliac\_vein\_right    						& 0.986$\pm$0.058    &       & vertebrae\_T10    	& 0.935$\pm$0.212    &       & rib\_left\_12    	& 0.884$\pm$0.277 \\
\cmidrule{2-6}\cmidrule{8-9}\cmidrule{11-12}          				& esophagus    				& 0.971$\pm$0.062    & \multirow{9}[18]{*}{G3} & gluteus\_medius\_right		& 0.968$\pm$0.110    &       & vertebrae\_T11    	& 0.940$\pm$0.204    &       & rib\_right\_4		& 0.658$\pm$0.413 \\
\cmidrule{2-3}\cmidrule{5-6}\cmidrule{8-9}\cmidrule{11-12}          & stomach    				& 0.963$\pm$0.063    &       & gluteus\_medius\_left    					& 0.997$\pm$0.014    &       & vertebrae\_T12    	& 0.936$\pm$0.219    &       & rib\_right\_5    	& 0.771$\pm$0.348 \\
\cmidrule{2-3}\cmidrule{5-6}\cmidrule{8-9}\cmidrule{11-12}          & pancreas   				& 0.962$\pm$0.080    &       & gluteus\_maximus\_left   					& 0.921$\pm$0.251    &       & vertebrae\_L1   	& 0.942$\pm$0.190   &       & rib\_right\_6  	& 0.817$\pm$0.315 \\
\cmidrule{2-3}\cmidrule{5-6}\cmidrule{8-9}\cmidrule{11-12}          & adrenal\_gland\_right   	& 0.879$\pm$0.212    &       & gluteus\_maximus\_right   					& 0.893$\pm$0.243    &       & vertebrae\_L2   	& 0.955$\pm$0.142   &       & rib\_right\_7   	& 0.845$\pm$0.288 \\
\cmidrule{2-3}\cmidrule{5-6}\cmidrule{8-9}\cmidrule{11-12}          & adrenal\_gland\_left   	& 0.846$\pm$0.235    &       & autochthon\_left   							& 0.999$\pm$0.001    &       & vertebrae\_L3   	& 0.957$\pm$0.169   &       & rib\_right\_8   	& 0.887$\pm$0.240 \\
\cmidrule{2-3}\cmidrule{5-6}\cmidrule{8-9}\cmidrule{11-12}          & duodenum    				& 0.942$\pm$0.065    &       & autochthon\_right   							& 0.999$\pm$0.001    &       & vertebrae\_L4   	& 0.944$\pm$0.197   &       & rib\_right\_9   	& 0.906$\pm$0.227 \\
\cmidrule{2-3}\cmidrule{5-6}\cmidrule{8-9}\cmidrule{11-12}          & small\_bowel   			& 0.978$\pm$0.030    &       & iliopsoas\_right   							& 0.996$\pm$0.013    &       & vertebrae\_L5   	& 0.923$\pm$0.217   &       & rib\_right\_10   	& 0.906$\pm$0.241 \\
\cmidrule{2-3}\cmidrule{5-6}\cmidrule{8-9}\cmidrule{11-12}          & colon   					& 0.977$\pm$0.032    &       & iliopsoas\_left   							& 0.992$\pm$0.043    &       & rib\_left\_4   	& 0.672$\pm$0.442   &       & rib\_right\_11   	& 0.905$\pm$0.223 \\
\cmidrule{2-3}\cmidrule{5-6}\cmidrule{8-9}\cmidrule{11-12}          & -   						& -   				 &       & -   											& -   				 &       & rib\_left\_5   	& 0.795$\pm$0.340   &       & rib\_right\_12   	& 0.893$\pm$0.249 \\
    \bottomrule
    \end{tabular}%
    \end{adjustbox}
  \label{tab:NSD_score}%
\end{table*}%

\begin{table*}[htbp]
  \centering
  \caption{DSC and NSD results of MRISegmentator and nnUNet-AMOS on the internal dataset, and the validation subset of the external AMOS22 dataset.}
    \begin{adjustbox}{max width=\textwidth}
    \begin{tabular}{|c|c|c|c|c|c|c|}
    \toprule
    \multirow{3}[6]{*}{} & \multicolumn{4}{c|}{The internal dataset} & \multicolumn{2}{c|}{The AMOS22 dataset} \\
	\cmidrule{2-7}          & \multicolumn{2}{c|}{MRISegmentator} & \multicolumn{2}{c|}{nnUnet-AMOS} & \multicolumn{2}{c|}{MRISegmentator} \\
	\cmidrule{2-7}          & DSC   				& NSD   				& DSC   				& NSD   				& DSC   			& NSD \\
    \midrule	
    Spleen 					&0.977$\pm$0.015       	& 0.982$\pm$0.019      	& 0.939$\pm$0.131		& 0.931$\pm$0.139      	& 0.967$\pm$0.024	& 0.981$\pm$0.034	\\
    \midrule
    Kidney\_right 			&0.984$\pm$0.010       	& 0.988$\pm$0.015      	& 0.967$\pm$0.091      	& 0.979$\pm$0.028      	& 0.963$\pm$0.020	& 0.974$\pm$0.027	\\
    \midrule
    Kidney\_left 			&0.970$\pm$0.093       	& 0.968$\pm$0.091      	& 0.962$\pm$0.114      	& 0.947$\pm$0.118      	& 0.951$\pm$0.077	& 0.966$\pm$0.069	\\
    \midrule
    Gallbladder 			&0.918$\pm$0.109      	& 0.969$\pm$0.069     	& 0.779$\pm$0.283      	& 0.852$\pm$0.264      	& 0.776$\pm$0.249	& 0.827$\pm$0.264 	\\
    \midrule
    Esophagus 				&0.915$\pm$0.107       	& 0.971$\pm$0.062      	& 0.815$\pm$0.148     	& 0.955$\pm$0.077      	& 0.681$\pm$0.193	& 0.866$\pm$0.180	\\
    \midrule
    Liver 					&0.993$\pm$0.002       	& 0.993$\pm$0.009      	& 0.984$\pm$0.019      	& 0.982$\pm$0.031      	& 0.975$\pm$0.009	& 0.966$\pm$0.024 	\\
    \midrule
    Stomach 				&0.965$\pm$0.044       	& 0.963$\pm$0.063      	& 0.837$\pm$0.199      	& 0.855$\pm$0.190      	& 0.877$\pm$0.136	& 0.879$\pm$0.143 	\\
    \midrule
    Aorta 					&0.990$\pm$0.006      	& 0.999$\pm$0.003     	& 0.960$\pm$0.031      	& 0.990$\pm$0.022      	& 0.947$\pm$0.043	& 0.991$\pm$0.032 	\\
    \midrule
    IVC   					&0.974$\pm$0.019		& 0.996$\pm$0.014     	& 0.902$\pm$0.070      	& 0.954$\pm$0.070      	& 0.878$\pm$0.069	& 0.959$\pm$0.055 	\\
    \midrule	
    Pancreas 				&0.931$\pm$0.080       	& 0.962$\pm$0.080      	& 0.841$\pm$0.174      	& 0.896$\pm$0.167      	& 0.853$\pm$0.105	& 0.913$\pm$0.112 	\\
    \midrule
    Adrenal gland\_right 	&0.669$\pm$0.203       	& 0.879$\pm$0.212      	& 0.544$\pm$0.231     	& 0.765$\pm$0.251       & 0.607$\pm$0.137	& 0.886$\pm$0.105 	\\
    \midrule
    Adrenal gland\_left 	&0.653$\pm$0.217       	& 0.846$\pm$0.235      	& 0.458$\pm$0.257      	& 0.660$\pm$0.306     	& 0.618$\pm$0.210	& 0.820$\pm$0.242 	\\
    \midrule
    Duodenum 				&0.898$\pm$0.073       	& 0.942$\pm$0.065      	& 0.774$\pm$0.156      	& 0.849$\pm$0.142      	& 0.685$\pm$0.134	& 0.783$\pm$0.143 	\\
    \bottomrule
    \end{tabular}%
    \end{adjustbox}
  \label{tab:NSD_mriseg_amos22}%
\end{table*}%

\newpage
\subsection{Example segmentation results of patients from the AMOS22 dataset and Duke Liver dataset}

\begin{figure}[!htb]
    \centering
    \begin{minipage}{1\textwidth}
        \centering
        \includegraphics[width=1\linewidth]{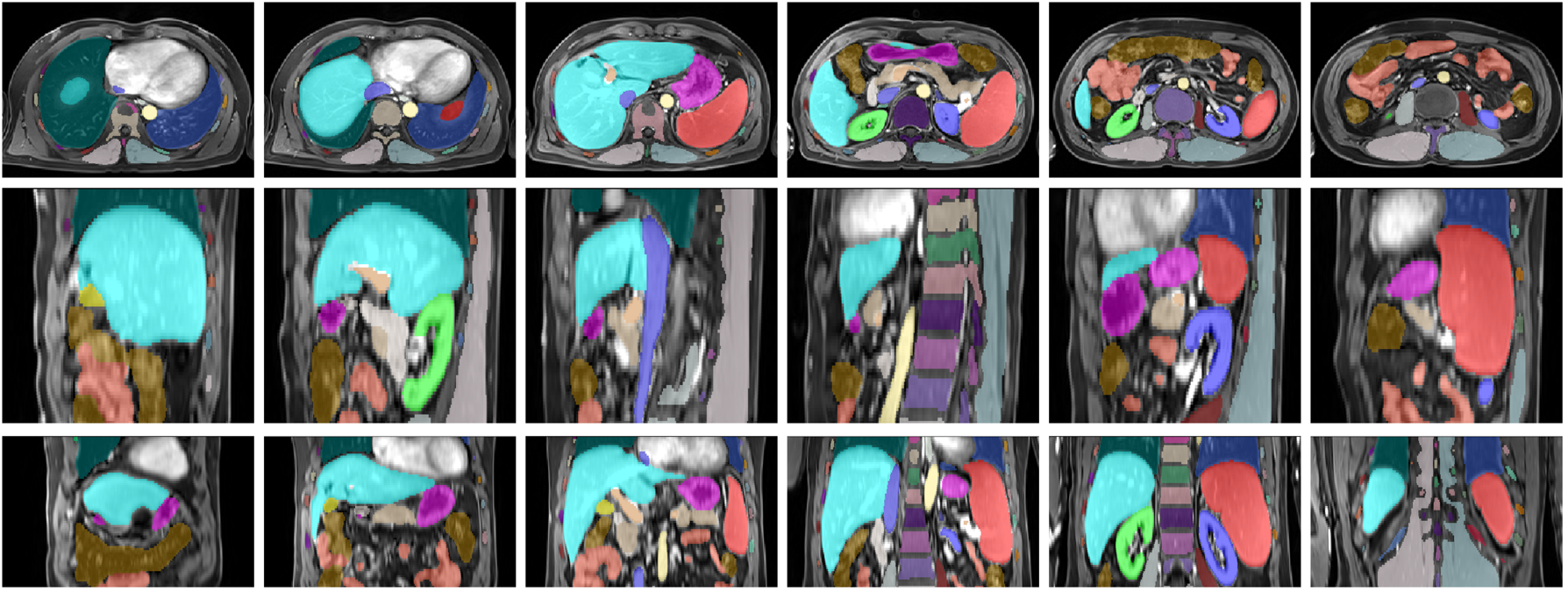} 
        \caption{Segmentation results of one patient from the AMOS dataset. Top, middle, and bottom rows show axial, sagittal, and coronal views.}
        \label{Fig:amos_result} 
    \end{minipage}
\end{figure}

\begin{figure}[!htb]
    \begin{minipage}{1\textwidth}
    \centering
    \includegraphics[width=1\linewidth]{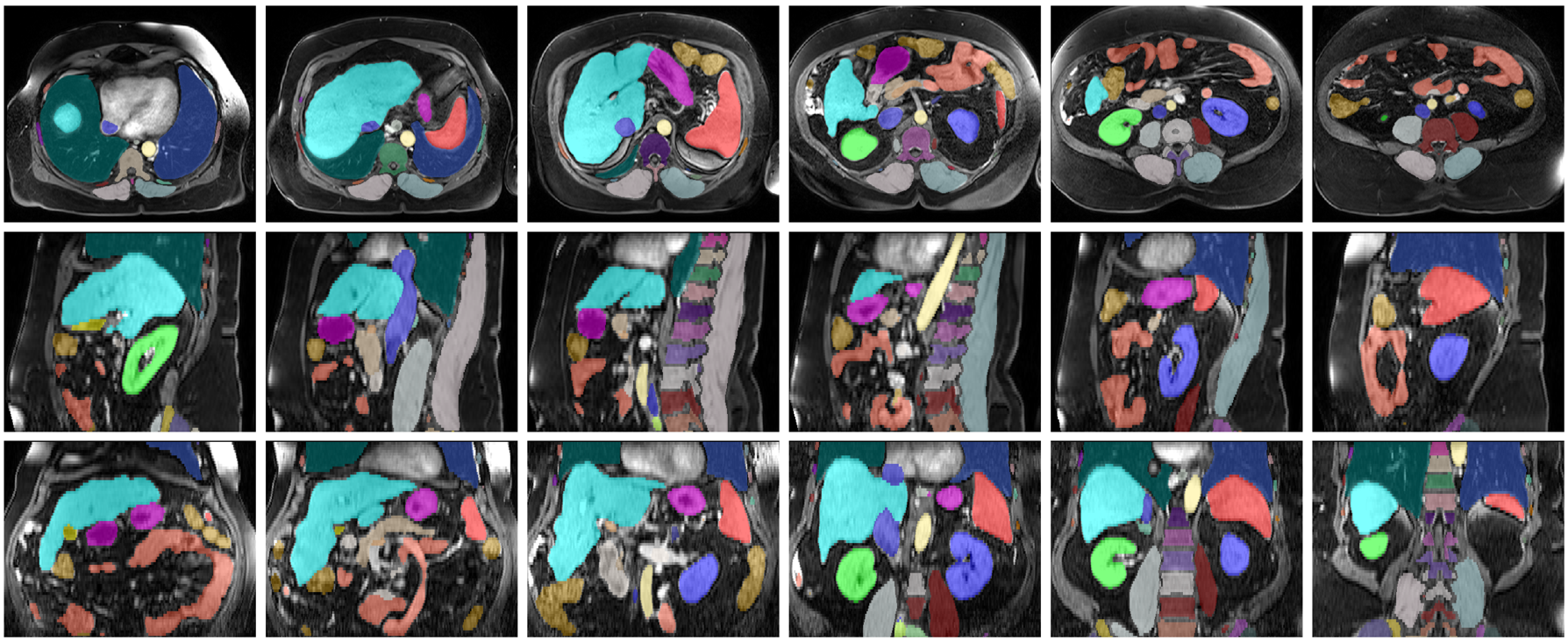} 
    \caption{Segmentation results of one patient from the Duke Liver dataset. Top, middle, and bottom rows show axial, sagittal, and coronal views.}
    \label{Fig:duke_result} 
    \end{minipage}
\end{figure}

Figs.~\ref{Fig:amos_result} and \ref{Fig:duke_result} show the segmentation results for the AMOS22 and Duke Liver datasets.

\begin{figure}[!htb]
    \begin{minipage}{1\textwidth}
        \centering
        \includegraphics[width=1\linewidth]{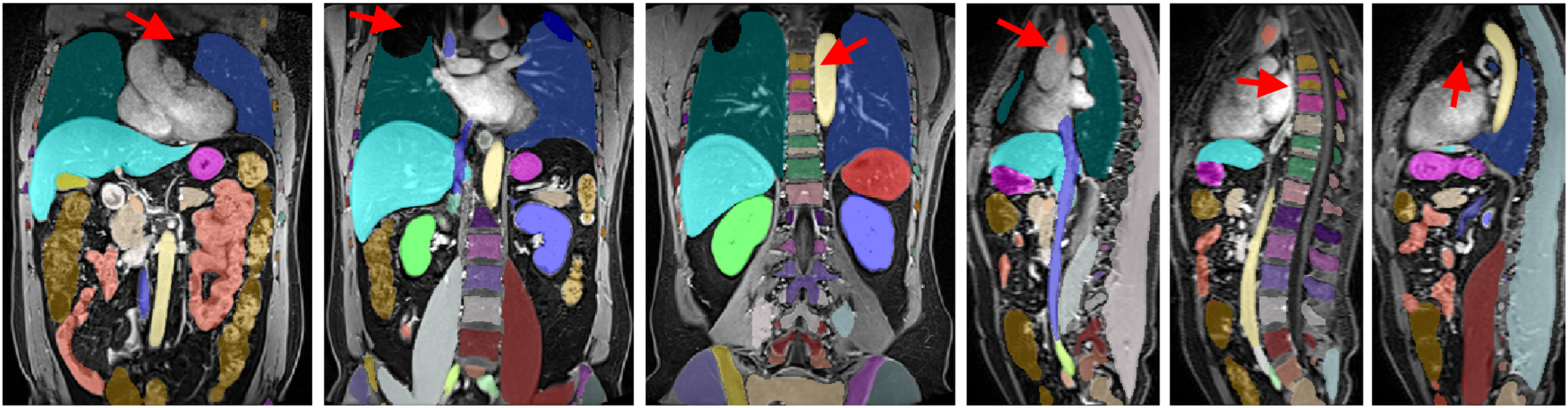} 
        \caption{Segmentation results of a patient in coronal-view T1w MRI from the AMOS22 dataset. The red arrows indicated the segmentation errors in the chest.}
        \label{Fig:amos_coronal_bad_example} 
    \end{minipage}
\end{figure}

\subsection{Color map of the segmentation mask}

\begin{figure}[!htb]
    \centering
    \begin{minipage}{1\textwidth}
        \centering
        \includegraphics[width=1\linewidth]{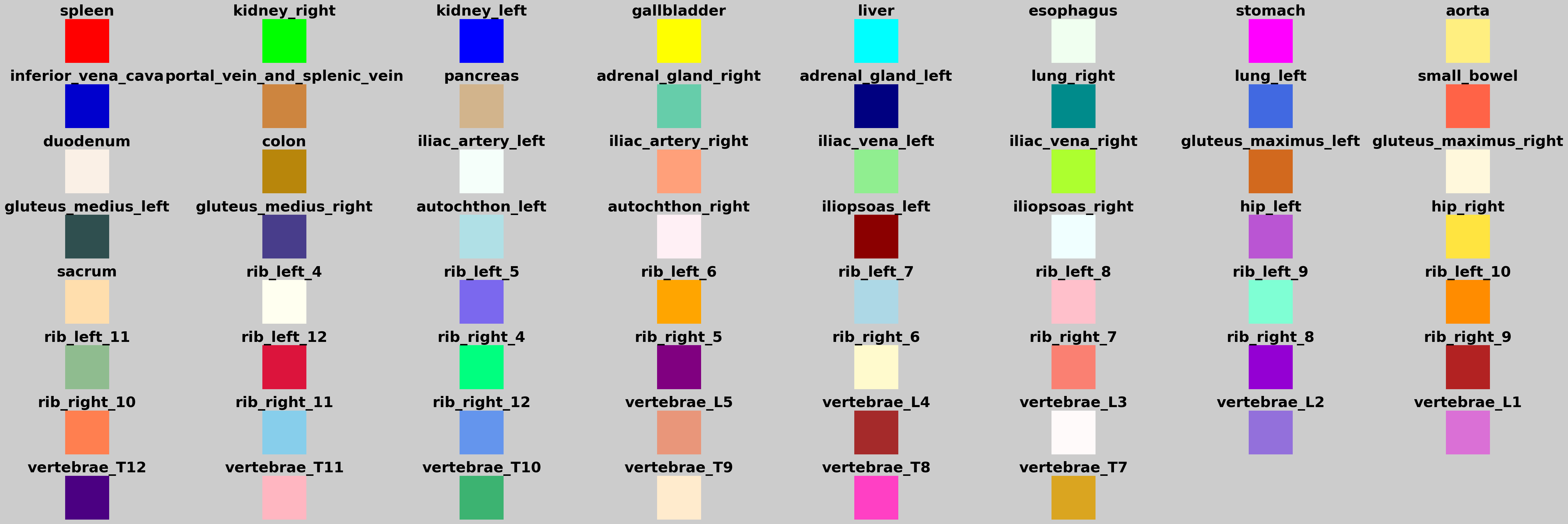} 
    \end{minipage}
\caption{The color map of the segmentation masks for the 62 organs and structures used for visualization.}
\label{Fig:organLegend} 
\end{figure}

\end{document}